\documentclass[aps,prl,superscriptaddress,showpacs,twocolumn,10pt]{revtex4-1}
\usepackage{graphicx}
\usepackage{amsmath}
\usepackage{dsfont}
\usepackage{xspace}
\usepackage{color}
\usepackage{fixmath}
\usepackage[normalem]{ulem}

\usepackage{siunitx}
\DeclareSIUnit\Er{\Er}

\DeclareMathOperator{\re}{\mbox{Re}}
\DeclareMathOperator{\im}{\mbox{Im}}

\newcommand{\ve}[1]{{\bf #1}}

\newcommand{\nag}{{\phantom{\dagger}}}

\newcommand{\eqw}[1]{(\ref{#1})}
\newcommand{\eq}[1]{Eq.\thinspace{}(\ref{#1})}

\newcommand{\fig}[1]{Fig.\thinspace{}\ref{#1}}

\newcommand{\fc}[1]{({#1})}

\newcommand{\down}{\ensuremath{\ket{\downarrow}\ }}
\newcommand{\up}{\ensuremath{\ket{\uparrow}\ }}

\renewcommand{\sp}{\chi\left(\ve{Q}\right)}

\def\ket#1{\mathinner{|{#1}\rangle}}
\def\braket#1{\mathinner{\langle{#1}\rangle}}

\makeatletter
\newsavebox{\@brx}
\newcommand{\llangle}[1][]{\savebox{\@brx}{\(\m@th{#1\langle}\)}%
  \mathopen{\copy\@brx\kern-0.5\wd\@brx\usebox{\@brx}}}
\newcommand{\rrangle}[1][]{\savebox{\@brx}{\(\m@th{#1\rangle}\)}%
  \mathclose{\copy\@brx\kern-0.5\wd\@brx\usebox{\@brx}}}
\makeatother

\begin{document}

\title{Far-from-equilibrium spin transport in Heisenberg quantum magnets}

\author{Sebastian Hild}%
\author{Takeshi Fukuhara}%
\altaffiliation[Present address: ]{RIKEN Center for Emergent Matter Science (CEMS), Wako, 351-0198, Japan}%
\author{Peter~Schau\ss}%
\author{Johannes~Zeiher}%
\affiliation{Max-Planck-Institut f\"{u}r Quantenoptik, 85748 Garching, Germany}
\author{Michael Knap}%
\affiliation{Department of Physics, Harvard University, Cambridge, Massachusetts 02138, USA}%
\affiliation{ITAMP, Harvard-Smithsonian Center for Astrophysics, Cambridge, Massachusetts 02138, USA}%
\author{Eugene Demler}%
\affiliation{Department of Physics, Harvard University, Cambridge, Massachusetts 02138, USA}%
\date{\today}
\author{Immanuel Bloch}%
\affiliation{Max-Planck-Institut f\"{u}r Quantenoptik, 85748 Garching, Germany}
\affiliation{Ludwig-Maximilians-Universit\"{a}t, Fakult\"{a}t f\"{u}r Physik, 80799 M\"{u}nchen, Germany}%
\author{Christian Gross}%
\affiliation{Max-Planck-Institut f\"{u}r Quantenoptik, 85748 Garching, Germany}

\begin{abstract}
  We study experimentally the far-from-equilibrium dynamics in ferromagnetic Heisenberg quantum magnets realized with ultracold atoms in an optical lattice. After controlled imprinting of a spin spiral pattern with adjustable wave vector, we measure the decay of the initial spin correlations through single-site resolved detection. On the experimentally accessible timescale of several exchange times we find a profound dependence of the decay rate on the wave vector. In one-dimensional systems we observe diffusion-like spin transport with a dimensionless diffusion coefficient of 0.22(1). We show how this behavior emerges from the microscopic properties of the closed quantum system. In contrast to the one-dimensional case, our transport measurements for two-dimensional Heisenberg systems indicate anomalous super-diffusion.
\end{abstract}


\maketitle

Since its introduction, the Heisenberg spin model has posed fundamental challenges for the understanding of non-equilibium dynamics in quantum magnets. On a very basic, phenomenological level, the concept of spin diffusion was introduced more than 60 years ago\,\cite{bloembergen_interaction_1949,van_hove_time-dependent_1954,de_gennes_inelastic_1958}. It has been commonly applied to interpret nuclear magnetic resonance spin lattice relaxation and electron spin resonance experiments~\cite{hone_proton_1974,boucher_high-temperature_1976,benner_experimental_1978,takigawa_dynamics_1996,thurber_o17_2001}. However, up to now it has never been justified {\it ab-inito} from a microscopic model. Moreover, many analytical and numerical studies suggested the existence of anomalous diffusion in Heisenberg models at high temperature, because of non-trivial commutation relations between spin operators leading to a failure of usual hydrodynamics~\cite{chertkov_long-time_1994,lovesey_time-dependent_1994,muller_anomalous_1988,de_alcantara_bonfim_breakdown_1992}. The strongest evidence for anomalous diffusion resulted from the memory function approach~\cite{chertkov_long-time_1994,lovesey_time-dependent_1994} and classical numerical simulations~\cite{muller_anomalous_1988,de_alcantara_bonfim_breakdown_1992}. In one dimension, Heisenberg models have the additional property of being integrable~\cite{bethe_zur_1931}. As a result, at zero temperature the linear spin response is ballistic in the gapless phase~\cite{shastry_twisted_1990} while at finite temperature no definite conclusion could be reached so far~\cite{castella_integrability_1995,zotos_transport_1997,zotos_finite_1999,narozhny_transport_1998,alvarez_low-temperature_2002,heidrich-meisner_zero-frequency_2003,sirker_diffusion_2009,sirker_conservation_2011,prosen_open_2011,znidaric_spin_2011,steinigeweg_spin_2011,karrasch_finite-temperature_2012,znidaric_anomalous_2014}. It has been argued that the regular, non-ballistic contribution to spin transport can indeed be of diffusive character at finite temperature~\cite{sirker_diffusion_2009,sirker_conservation_2011}.

\begin{figure} \centering \includegraphics[width=0.48\textwidth]{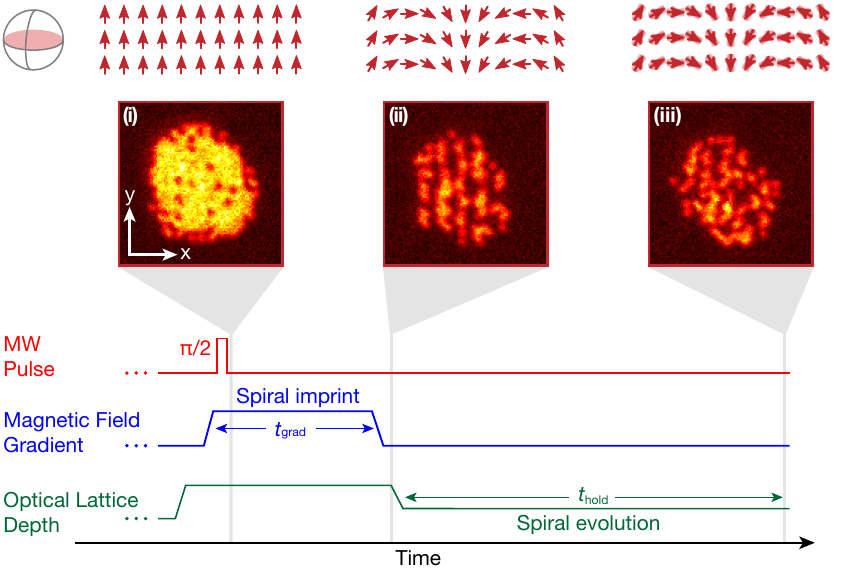}
  \caption{Experimental sequence for the measurement of the spiral evolution.
    The illustrations on top show the spin distribution in the transverse plane
    at different stages of the experiment as indicated by the gray shading
    (before (i) and after (ii) spiral imprinting and after evolution (iii)).
    The pictures below are single shot measurements at the respective
    times after an additional $\pi/2$-pulse (not shown) and removal of the $\up$
    state. The experimental sequence is depicted at the bottom.}
    \label{fig:expt}
\end{figure}

To address this fundamental problem, we experimentally study the far-from-equilibrium dynamics of quantum spins in one and two dimensions, realized with ultracold atoms in optical lattices. In our study, we prepare initial spin spiral states of a defined wave vector and track their relaxation dynamics. Our study is also motivated by recent experiments on spin diffusion in ultracold fermions~\cite{sommer_universal_2011,koschorreck_universal_2013,bardon_transverse_2014}, which found an exceptionally low transverse spin diffusion constant in two dimensions~\cite{koschorreck_universal_2013}, very different from three dimensional results~\cite{bardon_transverse_2014}. These so far unexplained results motivate studies in alternative systems to check the generality of the observation. In our experiment and numerical simulations, we find that spin dynamics at high-energy-density in one-dimensional Heisenberg systems exhibits diffusive character. An intuitive way to understand the emergence of such a classical-like transport is given through interaction induced dephasing between the many-body eigenstates spanning the initial spin spiral state. In contrast, the 2D system is shown to exhibit anomalous super-diffusion for the observed intermediate timescales, in agreement with earlier predictions~\cite{lovesey_time-dependent_1994}. We find in both cases that the closed quantum evolution at high-energy-density is in stark contrast to the one of a few excited magnons, which propagate ballistically~\cite{ganahl_observation_2012, fukuhara_quantum_2013, fukuhara_microscopic_2013}.

Following the concept of spin-grating spectroscopy~\cite{cameron_spin_1996,zhang_first_1998,wang_gate_2013}, we prepare initial large amplitude transverse spin spirals $\ket{\sp}=\prod_j  (\ket{\uparrow}_j +e^{-i \ve{Q} \cdot \ve{x}_j} \ket{\downarrow}_j)$ with a controlled wave vector $\ve{Q}$, where $\ve{x}_j$ is the position of the lattice sites. On a phenomenological level the evolution of the spiral would be captured through the dynamics of a single component $M_{\perp}$ of the transverse magnetization. Combination of the continuity equation and the empirical Fick's law leads to the diffusion equation $\partial M_\perp /\partial t = D \nabla^2 M_\perp$, with a diffusion constant $D$. This equation predicts a characteristic dependence of the lifetime $\tau$ of the transverse magnetization $M_\perp$ on the initial wave vector $1/\tau = D |\ve{Q}|^2$.
In order to test this prediction far from equilibrium, where a vast number of states are available to scatter, we track the relaxation dynamics of the spin spiral with single-site resolution and compare our experiments to numerical simulations.

\begin{figure}
  \centering
  \includegraphics{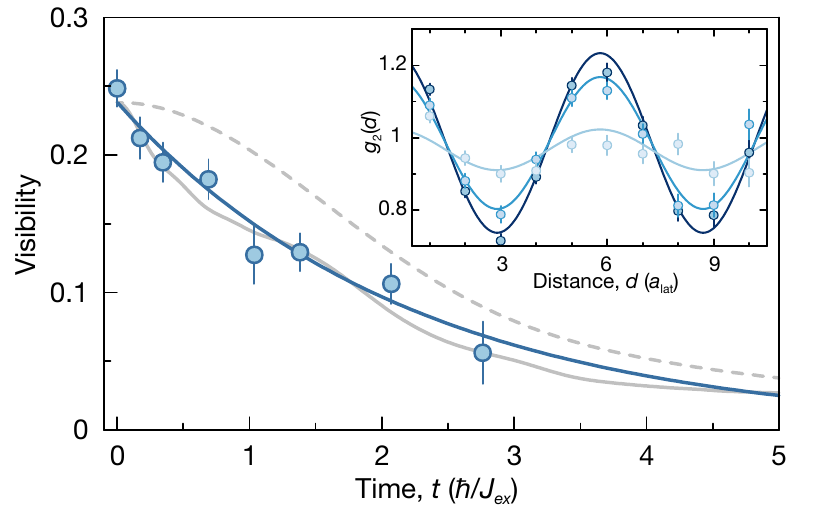}
  \caption{Decay of a 1D spin spiral. Measured decay of an exemplary 1D spin spiral with wavelength $\lambda=5.7\,a_\mathrm{lat}$ at $10\,E_r$ lattice depth. The solid blue line is an exponential fit used to extract the lifetime. We also show theoretical predictions of the Heisenberg model (gray, dashed line) and the $t$-$J$ model for $0.08$ hole probability (gray, solid line). The inset shows three measured correlations $g_2(d)$ at $t_1=0$, $t_2=0.7\,\hbar/J_{ex}$, $t_3=2.8\,\hbar/J_{ex}$ (dark to bright blue), from which the visibility is extracted via sinusoidal fits.} \label{fig:decay}
\end{figure}

We implement the spin Hamiltonian using ultracold bosonic $^{87}$Rb atoms in an optical lattice, initially prepared in a Mott insulating regime with unity filling. In this strong coupling regime, our atomic lattice system can be mapped to the ferromagnetic Heisenberg model~\cite{duan_controlling_2003, kuklov_counterflow_2003,altman_phase_2003} which is slightly modified in our case due to a small number of mobile particle-hole defects:
\begin{equation}
 \hat H = -J_{ex} \sum_{i}\left[\frac{1}{2}\left(\hat{{S}}^+_i \hat{{S}}^-_{i+1}+ \hat{{S}}^-_i \hat{{S}}^+_{i+1}\right) + \Delta \hat{{S}}^z_i \hat{{S}}^z_{i+1}\right] + \hat H_d\;.
 \label{eq:H}
\end{equation}
Here $J_{ex}\approx4J^2/U$ is the superexchange coupling, and $J$ and $U$ denote the hopping and interaction energy scales of the underlying single-band Hubbard model. We note that for the spin states employed in the experiment, the interaction energies for different spin channels vary only on the level of $1\,$\% resulting in an almost isotropic model with $\Delta \approx 1$~\cite{pertot_collinear_2010, sup_mat}. The spin operators  are defined through the boson creation and annihilation operators ${\hat b}^\dag_{\sigma,i}$ and $ {\hat b}^\nag_{\sigma,i} $ for the two spin states $\sigma = {\uparrow,\downarrow}$ as $\hat{S}^+_i={\hat b}_{\uparrow,i}^\dag {\hat b}^\nag_{\downarrow,i} $, $\hat{S}^-_i={\hat b}_{\downarrow,i}^\dag {\hat b}^\nag_{\uparrow,i} $ and $\hat{{S}}^z_i=\left(\hat{n}_{\uparrow,i}-\hat{n}_{\downarrow,i}\right)/2$. The last term, $\hat H_d$, in \eq{eq:H} describes the dynamics of defects. Here we restrict the discussion to the dominating effect of holes in the Mott insulator. The probability of doubly occupied sites
is assumed to be lower and thus neglected. The Hamiltonian in \eq{eq:H} then corresponds to the bosonic $t$-$J$ model~\cite{auerbach_interacting_1994}.

Our experiments started with the preparation of a degenerate $^{87}$Rb Bose gas confined in a single anti-node of a vertical optical standing wave. The two-dimensional gas was then driven into the Mott insulating phase with unity filling by adiabatically switching on a horizontal square lattice with lattice spacing $a_{\text{lat}}=532$\,nm. Two long-lived hyperfine states ($\down\equiv\ket{F=1,m_{F}=-1}$ and $\up \equiv \ket{2,-2}$) are used as a pseudo spin-$1/2$ system. For the preparation of the initial spiral, all many-body spin dynamics was suppressed in a deep optical lattice of $20\,E_r$ lattice depth. Here $E_r=h^2 /(8m a_\mathrm{lat}^2)$ denotes the recoil energy of the lattice, with $m$ being the atomic mass. A global $\pi/2$-pulse of \SI{10}{\micro\second} duration then transferred all atoms to a symmetric superposition of the two hyperfine states. Next, a relative phase between neighboring spins was imprinted by exposing the atoms to a constant magnetic field gradient of 0.2\,G/cm (corresponding to a frequency shift of 20\,Hz/$a_\text{lat}$). Time evolution in the gradient field leads to a linear growth of this relative phase over time and thus imprints a controlled spin spiral state $|\chi(\ve{Q})\rangle$. Subsequently, the gradient was reduced to a negligible value of $\leq 2$\,mG/cm for the further course of the experiment. The evolution of the strongly-interacting spins was then initiated by lowering the depth of either one or both of the horizontal lattices within 5\,ms to the desired value between 8-16\,$E_r$ for the experiments in 1D or 2D, respectively. The experiments in 1D were carried out in weaker lattices as the transition point towards the superfluid region occurs at lower lattice depth. For the ramp-down, we chose a timescale that both minimizes heating, while still being short compared to the ensuing spin dynamics. Then the system was let to evolve for variable times of up to $t_{\text{hold}}\simeq 3\hbar/J_{ex}$. For detection, the final spin configuration was frozen by rapidly increasing the lattice depth within 1\,ms to 40\,$E_r$. A second $\pi$/2-pulse then completed the global Ramsey interferometer by rotating the transverse spiral to the measurement basis along the $z$-direction. Finally, the $\up$ state was optically removed from the lattice and the remaining atoms $n_{j}$ per site $j$ in the $\down$ component were imaged with single-site resolved fluorescence detection \cite{sherson_single-atom-resolved_2010} (see Fig.~\ref{fig:expt}).

We analyze the spiral pattern through a second order correlation function $g_2(j,k)=\langle n_{j} n_{k}\rangle/\left( \langle n_{j} \rangle \langle n_{k}\rangle \right)$ and thereby avoid cancellation of the spiral signal due to shot-to-shot fluctuations in its phase caused by uniform magnetic field fluctuations. Note that in this case $g_2\left(j,k\right)$ is equivalent to $\re\, \langle \hat S^+_{j} \hat S^-_{k}\rangle$ when neglecting defects\,\cite{sup_mat}. The correlation signal dominantly depends on the distance $d$ between sites, such that $g_2\left(d\right)=N\sum_j \langle n_{j}n_{j+d}\rangle /( \sum_j \langle n_{j}\rangle )^2$ can be used to improve the signal-to-noise ratio. Here $\langle \cdot \rangle$ represents the ensemble average over different experimental realizations, whereas the sum describes the spatial average over $N$ different positions.

For increasing times $t_{\text{hold}}$, we observe a decay of the visibility of the spiral pattern, while its period remains unchanged. An exemplary dataset for such a dynamics in 1D is shown in Fig.\,\ref{fig:decay} for an initial spiral with wavelength $\lambda = 5.7(1)a_{\textrm{lat}}$. From an exponential fit to the decaying visibility, we find a lifetime of $\tau=30(3)$\,ms corresponding to $2.1(3)\,\hbar/J_\text{ex}$. We note that a simple mean-field treatment of the relaxational dynamics in the Heisenberg model based on a Landau-Lifshitz type evolution equation does not exhibit any dynamical evolution. Thus, quantum fluctuations beyond linear order are responsible for the decay of the spin spiral. We compare the experimentally observed decay to exact diagonalization predictions for the Heisenberg and the $t$-$J$ model taking the non-linearities fully into account (see Fig.\,\ref{fig:decay}). Both models predict an initial quadratic decay due to dephasing that happens on the fastest timescale ($\hbar/J_\mathrm{ex}$ or $\hbar/J$)\,\cite{sup_mat}. Experimentally, we only sample the decay on the superexchange timescale $\hbar/J_{ex}$ and thus cannot resolve the fast initial dynamics in the $t$-$J$ model.  While both models show good qualitative agreement with the experimental data, the $t$-$J$ model reproduces the observations for an independently characterized hole probability of $0.08(1)$.

\begin{figure}[h!]
  \centering
  \includegraphics{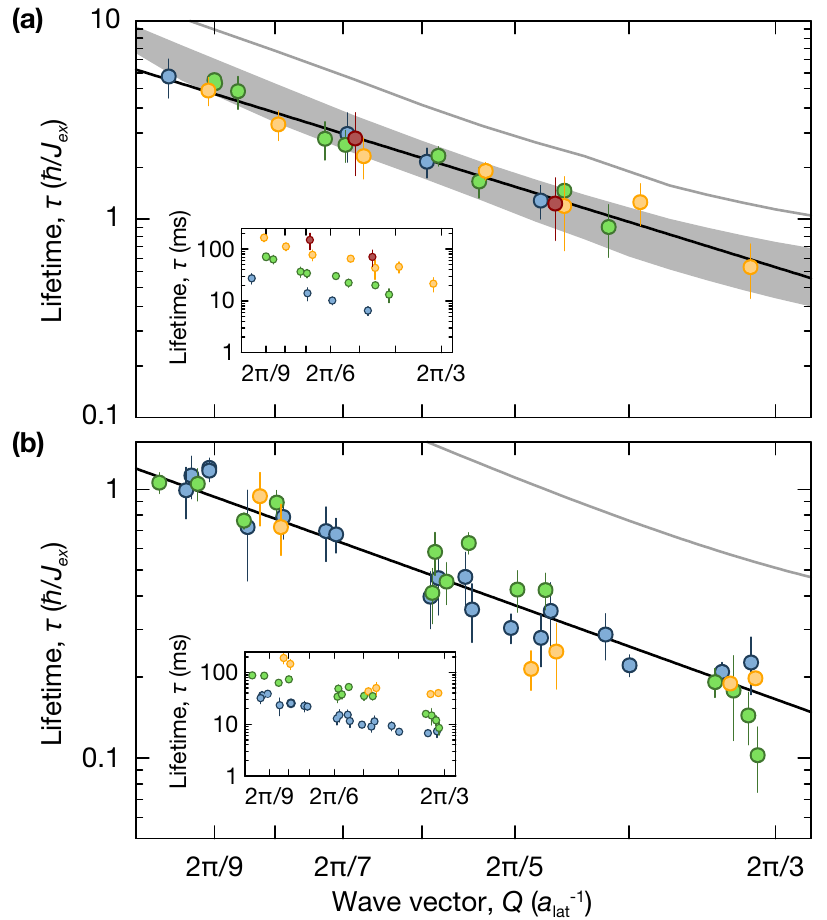}
  \caption{Wave vector dependence of the spin spiral lifetimes. We plot the data for 1D \fc{a} and 2D \fc{b} spirals double logarithmically and extract the exponents via power law fits (black lines). The lifetime is scaled with the superexchange rate $\hbar/J_\mathrm{ex}$, which results in a collapse of the measurements at different lattice heights [in 1D: $8\,E_r$ (blue), $10\,E_r$ (green), $12\,E_r$ (yellow), $13\,E_r$ (red) and in 2D:   $12\,E_r$ (blue),  $14\,E_r$ (green),  $16\,E_r$ (yellow)]. Additionally, predictions of the Heisenberg model (numerically solved in 1D, spin wave calculations in 2D) are shown as gray solid lines. The gray band in \fc{a} is obtained numerically from the $t$-$J$ model with hole probabilities between $0.04$ and $0.12$. The insets show the experimental data without scaling of the lifetimes.}  \label{fig:lifetime}
\end{figure}

In order to check the assumption of diffusion-like spin transport, we measure the lifetime $\tau$ for different wave vectors $\mathbf Q$, both in 1D and 2D. In 2D the spiral wave vectors were oriented diagonally to the lattice $\ve{Q}=(Q,Q)/\sqrt{2}$. The resulting data are shown in \fig{fig:lifetime} for both dimensionalities, different lattice depths and different initial wave vectors. When scaling the data with the exchange coupling $J_\text{ex}$, we find the datasets for different lattice depths to collapse. From this we deduce that $\hbar/J_{ex}$ is the relevant timescale for the main features of the observed dynamics and superexchange-mediated quantum magnetism is the underlying mechanism driving the dynamics.  In order to gain further insight into the wave vector dependence of the spiral lifetime, we plot the data in a
double-logarithmic plot and fit a power law with variable exponent $\alpha$ to the data $\tau \propto Q^{-\alpha}$. For our 1D data we find an exponent of $\alpha = 1.9(1)$ in good agreement with diffusive spin transport. In 2D the fitted exponent yields $\alpha = 1.6(1)$, differing notably from the one of diffusive transport and hinting at anomalous super-diffusion. For the analysis of the data, the exchange coupling $J_\text{ex}$ was extracted independently from single magnon propagation measurements following our earlier results in Ref.~\cite{fukuhara_quantum_2013}. In these measurements, we consistently find that the measured $J_{ex}$ is $20(10)$\% larger than the one calculated from {\it ab-initio} single-band calculations. We attribute this difference to interaction induced multi-band effects that are expected to effectively lower $U$, but raise $J$ \cite{Will_2010}.

\begin{figure}
  \centering
  \includegraphics{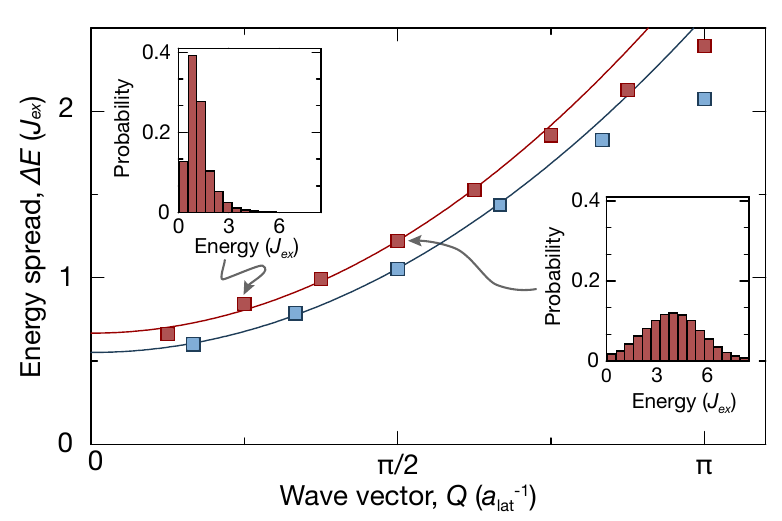}
  \caption{Microscopic view of the diffusion-like behavior in 1D. The energy spread of eigenstates contributing to the spin spiral decay grows quadratically with wave vector $Q$. This leads to the observed quadratic $Q$ dependence of the decay rate, as expected for classical spin diffusion. The data shown for two system sizes of $12$ (blue) and $16$ (red) sites is obtained from full diagonalization of 1D Heisenberg chains. Insets show energy histograms weighted with the overlap of the initial spiral and the eigenstates for $Q=\pi/4 a_\textrm{lat}$ and $Q=\pi/2 a_\textrm{lat}$ for systems with $16$ spins. A spin spiral with wave vector $Q$ is a superposition of many-body eigenstates with wave vectors $k$ that are integer multiples of $Q$.
} \label{fig:spectrum}
\end{figure}

The observed diffusion-like behavior can be understood microscopically in the one-dimensional case, where the numerical simulations based on the Heisenberg model also point to an approximately quadratic dependence of the decay rate on the wave vector in the experimentally accessible region. As the spiral state is not an eigenstate of the Heisenberg model, it shows overlap with several many-body eigenstates. Our simulations show that the energy spread $\Delta E$ in the many-body spectrum in fact increases quadratically with the spiral wave vector $Q$ (see Fig.~\ref{fig:spectrum}). The diffusive-like behavior in the evolution of the spiral state can thus be traced back to a many-body dephasing effect, with the shortest decay time occurring for a classical N\'eel state $Q=\pi/a_{\textrm{lat}}$ (see Ref.\,\cite{barmettler_relaxation_2009}).

When comparing the prediction in detail to the experimentally measured lifetimes (see Fig.\,\ref{fig:lifetime}), we find the latter to be shifted systematically to lower values. This behavior can be reproduced when considering the $t$-$J$ model with the measured hole probability, indicating a good qualitative and quantitative understanding of the evolution. The observations in the 2D situation are compared to results from a spin-wave theory prediction for the case without holes\,\cite{sup_mat}. While we find a similar qualitative behavior in this analysis, our experimental results are again shifted systematically towards lower lifetime values. Unfortunately, numerical simulations in 2D including holes remain currently out of reach because of the prohibitively large underlying Hilbert space.

The timescale of the diffusive behavior in 1D is set by the diffusion constant $D$. From dimensional analysis we find its natural units to be $\hbar/\widetilde{m}$, where $\widetilde{m}=\hbar^2/(2 J_\mathrm{ex} a_\mathrm{lat}^2)$ is the effective magnon mass. When assuming diffusive behavior (fixing the exponent $\alpha = 2$) we extract $D = 0.22 (1)\,\hbar/\widetilde{m}$ from our data. Remarkably, this is among the lowest values measured to date in a 1D many-body setting even though our measurements are carried out far from equilibrium in the highly excited regime.

\begin{figure}
  \centering
  \includegraphics{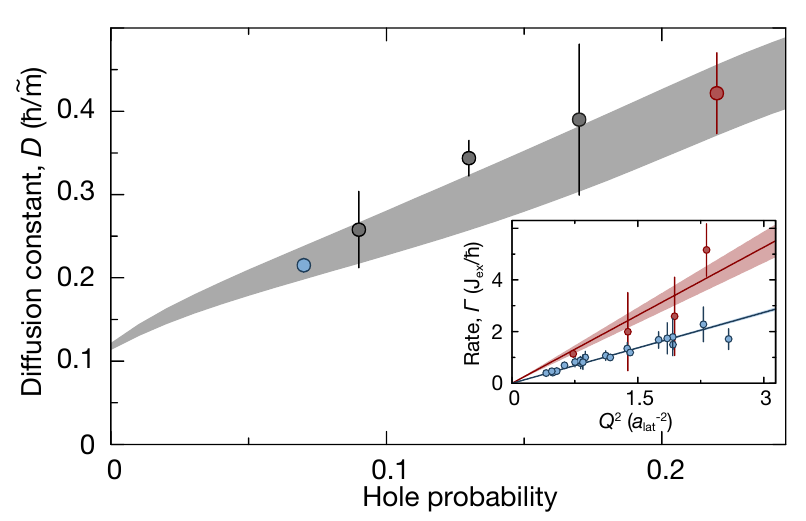}
  \caption{Dependence of the diffusion constant in 1D on the hole density. The diffusion constant $D$ increases approximately linearly with hole probability. The gray area is the numerical result of the $t$-$J$ model with its \SI{95}{\percent} confidence interval. The inset shows the decay rate $1/\tau$ of the spin spiral versus the squared wave vector $Q^2$ for the lowest (blue) and highest (red) hole probability.} \label{fig:holeprob}
\end{figure}

An intriguing additional question is the dependence of the diffusion constant on the hole density. In Fig.~\ref{fig:holeprob} we compare all 1D measurements for the lowest possible hole probability (the same data as shown in Fig.~\ref{fig:lifetime}) with data obtained for larger hole probabilities. Our data shows a clear trend towards increasing diffusion constant with increasing hole probability, consistent with numerical predictions based on the $t$-$J$ model. A linear increase can be indeed expected in 1D as each hole -- localized during the preparation -- introduces a fixed phase defect.

In conclusion, we have studied far-from-equilibrium spin transport in the Heisenberg model using high-energy-density spin spiral states in 1D and 2D. A numerical analysis explained the observed diffusion-like behavior in integrable 1D chains on a microscopic level. We found that the main features of the magnetic spin transport are robust against a small number of mobile hole defects in the system. In contrast to the diffusive behavior in 1D we observed anomalous super-diffusion in 2D Heisenberg magnets where integrability is broken. For future studies it would be interesting to explore long-time behavior which might in 1D shed light on the question of a residual ballistic transport\,\cite{castella_integrability_1995,zotos_transport_1997,zotos_finite_1999,narozhny_transport_1998,alvarez_low-temperature_2002,heidrich-meisner_zero-frequency_2003,sirker_diffusion_2009,sirker_conservation_2011,prosen_open_2011,znidaric_spin_2011,steinigeweg_spin_2011,karrasch_finite-temperature_2012,znidaric_anomalous_2014}, while in 2D it could unveil a possible crossover from a super-diffusive behavior to sub-diffusive behavior \cite{lovesey_time-dependent_1994}. Especially in 1D, it would be valuable to study spirals prepared with a wave vector close to $Q\sim \pi/a_\text{lat}$, where a transformation to the antiferromagnetic Heisenberg Hamiltonian is possible. Thus one can expect that the dynamics can be described with a Luttinger liquid formalism and predictions of Ref.~\cite{sirker_diffusion_2009,sirker_conservation_2011} could be tested. Furthermore, it would be interesting to study the absence of transport in interacting, many-body localized spin systems subject to quenched disorder~\cite{gornyi_interacting_2005,basko_metalinsulator_2006,oganesyan_localization_2007,pal_many-body_2010,bardarson_unbounded_2012,huse_phenomenology_2013,vosk_many-body_2013,serbyn_local_2013} using for instance local interferometric techniques~\cite{knap_probing_2013,serbyn_interferometric_2014}.

\begin{acknowledgments}
  We thank M. Cheneau, F. Heidrich-Meissner, T. Giamarchi, J. Thywissen, I. Affleck, M. Lukin and A. L\"auchli  for valuable discussions. The authors acknowledge support from MPG, EU (UQUAM), Harvard-MIT CUA, ARO-MURI Quism program, ARO-MURI on Atomtronics, as well as the Austrian Science Fund (FWF) Project No. J 3361-N20.
\end{acknowledgments}

\bibliography{References}

\newpage

\onecolumngrid
\newpage

\appendix

\begin{center}
{\large { \textbf{Supplemental Material for\\Far-from-equilibrium spin transport in Heisenberg quantum magnets} }}
\end{center}

\section{Experimental parameters}
  Former experiments have shown that the spin exchange coupling $J_{ex}$ is slightly stronger than expected from {\it ab-initio} single band estimations\,\cite{fukuhara_quantum_2013}. This difference can be attributed to multi-band effects which are expected to effectively lower $U$ but raise $J$\,\cite{Will_2010}. From previous experiments we deduce that {\it ab-initio} $J_{ex}$ values need to be correct by $20(10)$\%. In table~\ref{tab:parameter} we show a summary of the experimental parameters for the 1D case.

\begin{table}[h]
\begin{tabular}{c|c|c|c|c|c|c|c}
$V$ [Er] & $U/J$  & $J/\hbar$ [1/s] & $U/h$ [Hz] & $J_{ex}=4J^2/(U\hbar)$ [Hz] & $\hbar/J_{ex}$ [ms] & $J_{ex}'$ [Hz] & $\hbar/J_{ex}'$ [ms]  \\ \hline
8          & 9.8  & 392.3            & 614.7        & 159.4                                        & 6.3                & 191.3                    & 5.2                 \\
10         & 17.0 & 244.2            & 660.1        & 57.5                                         & 17.4              & 69.0                     & 14.5                 \\
12         & 28.1 & 156.0            & 698.3        & 22.2                                         & 45.1               & 26.6                     & 37.6                 \\
13         & 35.7 & 125.7            & 715.3        & 14.1                                         & 71.1               & 16.9                     & 59.2
\end{tabular}
\caption{Summary of parameters for the experiments in 1D.}
\label{tab:parameter}
\end{table}

\section{Defects in the Heisenberg model}

In the strongly interacting limit $J/U\ll1$, the dynamics of our system, which is initially prepared in a Mott insulating state with a certain hole probability, is described by the Hamiltonian

\begin{equation}
 \hat H = -\frac{J_{ex}}{2}\sum_{\langle i,j\rangle} \left(\hat S_i^+  \hat S_j^- + \Delta \hat S_i^z \hat S_j^z\right) + \hat H_d\;.
 \label{eq:Hfull}
\end{equation}
The first term is the Heisenberg Hamiltonian. In the strong coupling limit the parameters of this Hamiltonian are given by $J_{ex} = 4 J_\downarrow J_\uparrow/V$ and $J_{ex}\Delta = 4 J_\downarrow^2/U + 4 J_\uparrow^2/U - 2(J_\downarrow^2+J_\uparrow^2)/V$, where $U$ is the intraspecies and $V$ the interspecies interaction\,\cite{kuklov_counterflow_2003,duan_controlling_2003,altman_phase_2003}. While in principle tunable via a Feshbach resonance of ${}^{87}$Rb, in our experiment the parameters are chosen to be isotropic $J_\downarrow\sim J_\uparrow\sim J$ and $U\sim V$, leading to $\Delta \sim 1$. More precisely, the value of $\Delta = 0.986$ taking into account the slight anisotropies of the intra- and interspecies scattering length~\cite{pertot_collinear_2010}. The second term in \eqw{eq:Hfull} corresponds to a $tJ$-model that describes the dynamics of the holes for which $J_{ex}\gg J$ as $J/U$ is small and is explicitly given by
\begin{align}
 \hat H_d = - \sum_{\sigma, \langle i,j \rangle} J_\sigma^\nag \hat b_{\sigma i}^\dag \hat b_{\sigma j}^\nag
  -\sum_{\sigma, \langle i,j,k \rangle} \frac{J_\sigma^2}{V} \hat b_{\sigma i}^\dag \hat n_{\bar\sigma,j}^\nag \hat b_{\sigma k}^\nag
  + \frac{J_\uparrow J_\downarrow}{V} \hat b_{\bar\sigma i}^\dag  \hat S^\sigma_{j} \hat b_{\sigma k}^\nag
  +\frac{2 J_\sigma^2}{U} \hat b_{\sigma i}^\dag \hat n_{\sigma,j}^\nag \hat b_{\sigma k}^\nag\;,
\end{align}
where the second sum goes over nearest neighbor pairs $i,j$ and $j,k$ with $i\neq k$, $\bar \sigma$ flips the spin,
and $\hat S^\sigma_j$ is defined as $\hat S^+_j$ $(\hat S^-_j)$ for $\sigma={\uparrow\,(\downarrow)}$.

\begin{figure}[b]
        \centering
        \includegraphics[width=0.8\textwidth]{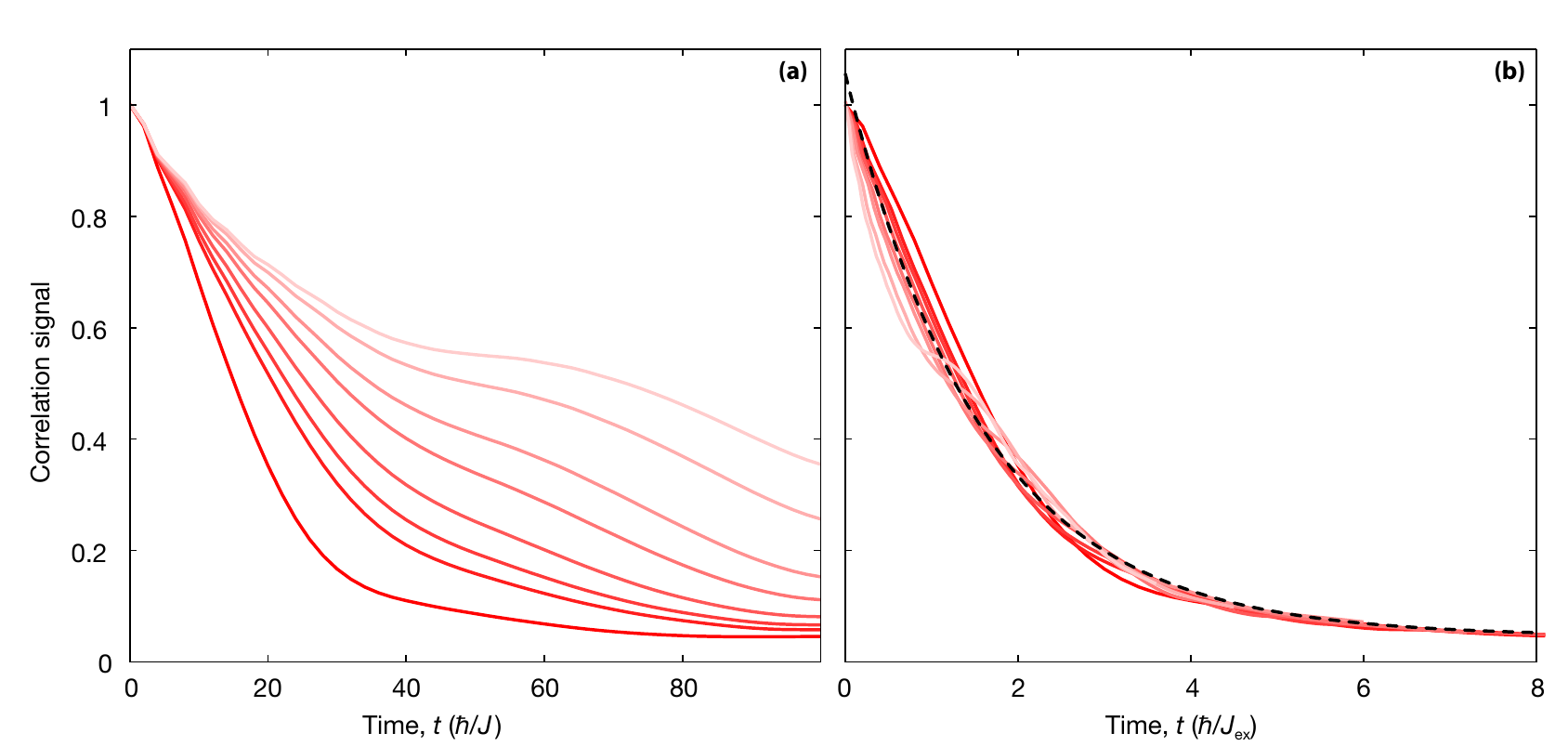}
        \caption{ Scaling of the spiral decay with finite hole probability.
        The decay of the spin spiral amplitude for a 1D ferromagnetic chain with 8\% hole probability is shown for various values of the lattice depth. In \fc{a} time $t$ is scaled by the bare kinetic energy of the bosons $J$ while in \fc{b} it is scaled by the superexchange interaction $J_{ex}$. In the latter case the data collapses onto a single curve for intermediate to long times, indicating that this decay is governed by superexchange processes. Dark to light red for $U=\lbrace \textrm{10, 15, 17, 20, 25, 30, 40, 50} \rbrace J$ and exponential fit in dashed black.}
        \label{fig:tJScaling}
\end{figure}

Using the full Hamiltonian \eqw{eq:Hfull} we simulate the spiral decay for finite hole probabilities. Even in the presence of holes, the spiral dynamics in intermediate to long times is determined by the Heisenberg superexchange interaction. This we numerically demonstrate in \fig{fig:tJScaling} which shows the spiral decay for a constant hole probability of $8\%$ but for different values of the lattice depth, encoded in $J/U$. For all values of the lattice depth, the short time dynamics is similar provided time is scaled by the kinetic energy of the bare bosons $J$ (see Fig.\,\ref{fig:tJScaling}a), while at intermediate to long times the decay is vastly different. In contrast, when the time is rescaled by the superexchange interaction $J_{ex}$ the intermediate to long time dynamics collapses for all $J/U$ (see Fig.\,\ref{fig:tJScaling}b). This indicates that the holes are distributed over the system on very short timescales determined by the boson kinetic energy. After this initial dynamics, the spiral decay is governed by the Heisenberg superexchange interactions.

\section{Spin-wave analysis of the spiral decay}

Another way of understanding the decay of the spin spiral states is to analyze the stability of their collective modes. Thus, we use spin-wave theory to identify the unstable modes and to predict the lifetime of the spiral.
We consider a system described by the Heisenberg Hamiltonian without defects and the initial state prepared in a spin spiral with wave vector $\ve{Q}$
\begin{equation}
 \langle S^+_j \rangle = S_0 e^{i\ve{Q}\,\ve{x}_j} \quad \langle S_j^z\rangle =0 \;.
\end{equation}
Here $S_0$ is the inverse length of the spin which is kept as a parameter in the following discussion. For our system of spins with two states $S_0=1/2$.
The spiral pattern breaks the translation symmetry. However, translational invariance can be restored by a local transformation into the twisted frame of the spiral
\begin{equation}
 \begin{pmatrix}
  S_i^x\\S_i^y\\S_i^z
 \end{pmatrix}=
 \begin{pmatrix}
  -\sin \ve{Q}\,\ve{x}_i & 0 & \cos \ve{Q}\,\ve{x}_i \\
  \cos \ve{Q}\,\ve{x}_i & 0 & \sin \ve{Q}\,\ve{x}_i \\
  0 & 1 & 0 \\
 \end{pmatrix}
 \begin{pmatrix}
  T_i^x\\T_i^y\\T_i^z
 \end{pmatrix}\;,
\end{equation}
leading to $\langle T_i^z \rangle=S_0$ and $\langle T_i^x \rangle=\langle T_i^y \rangle=0$. Furthermore the operators $T^\alpha_i$ obey Pauli spin algebra. In the transformed frame the Hamiltonian reads
\begin{equation}
 \hat H = -J \sum_{i,j} \cos \ve{Q}(\ve{x}_i-\ve{x}_j) (T_i^z T_{j}^z + T_i^x T_j^x) + \sin \ve{Q}(\ve{x}_i-\ve{x}_j) (T_i^z T_{j}^x - T_i^x T_j^z) + T_i^y T_j^y \;,
\end{equation}
which corresponds to an effective Dzyaloshinsky-Moriya interaction manifesting in the anisotropic spin exchange. Such Hamiltonians arise for instance in spin-orbit coupled systems. We employ a Holstein-Primakoff (HP) transformation
\begin{equation}
 T^z = S_0-a^\dag a^\nag \quad T^+=({2S_0-\hat n})^\frac{1}{2} a \quad T^-=a^\dag ({2S_0-\hat n})^\frac{1}{2}\;,
\end{equation}
which yields to quadratic order in the HP bosons
\begin{align}
 \hat H = &-2JS_0^2 N \sum_\alpha \cos Q_\alpha \nonumber\\
 &- J S_0 \sum_{\langle i,j \rangle} \cos \ve{Q} (\ve{x}_i-\ve{x}_j) \left[ \frac{1}{2} (a_i^\dag+a_i^\nag)(a_j^\dag+a_j^\nag)
 - (a_i^\dag a_i^\nag+a_j^\dag a_j^\nag)\right]
 -\frac{1}{2} (a_i^\dag-a_i^\nag)(a_j^\dag-a_j^\nag)+ O(\sqrt{S_0})\;.
\end{align}
A Bogoliubov transformation diagonalizes the Hamiltonian
\begin{equation}
 \hat H=\sum_{k} E_k \alpha_k^\dag \alpha_k^\nag\;, \qquad E_k=\sqrt{\epsilon_k^2 -\Delta_k^2}
\end{equation}
with $\epsilon_k = -JS \left[\sum_\alpha (\cos Q_\alpha+1)\cos k_\alpha - 2 \cos Q_\alpha\right]$ and $\Delta_k = JS \left[\sum_\alpha (\cos Q_\alpha-1)\cos k_\alpha\right]$.

\begin{figure}
        \centering
        \includegraphics[width=0.9\textwidth]{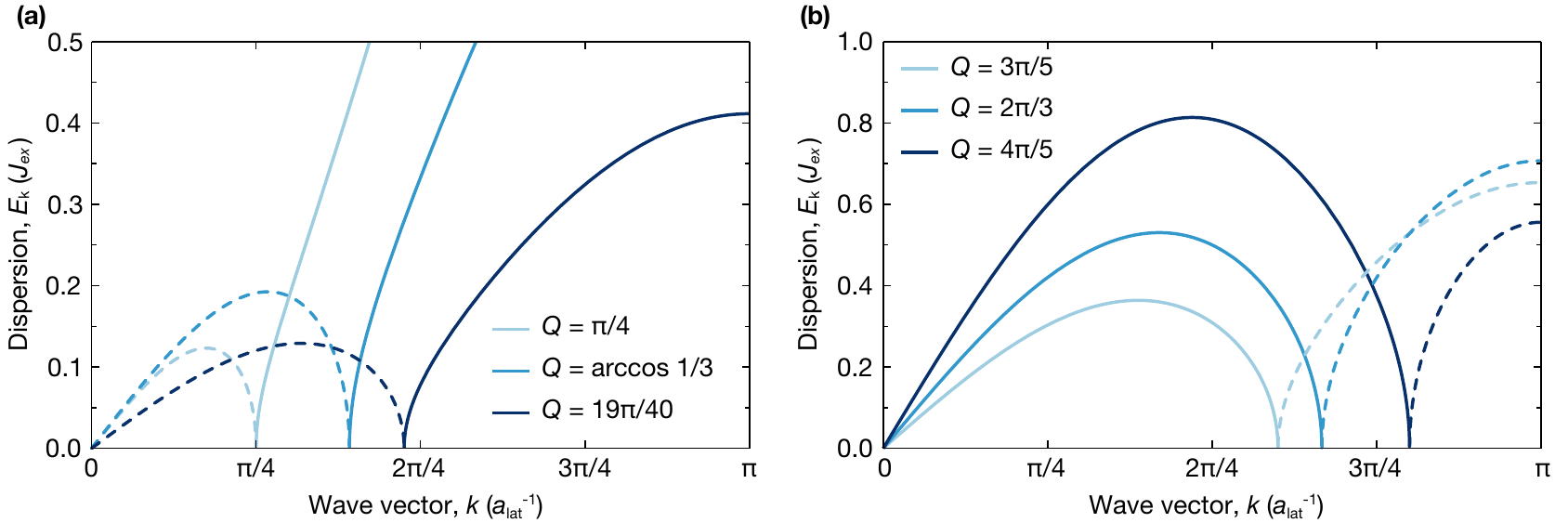}
        \caption{Real part (solid lines) and imaginary part (dashed lines) of the dispersion $E_k$ of a spin spiral with wave vector \fc{a} $Q<\pi/2a_\textrm{lat}$ 
        and \fc{b} $Q>\pi/2a_\textrm{lat}$. 
        Unstable modes are characterized by $\im E_k>0$.}
        \label{fig:unstablemodes}
\end{figure}

\textbf{Unstable modes.---}The dispersion $E_k$ can be imaginary for certain values of the wave vector $k$ leading to unstable modes, which when excited decay in time with a rate $\sim \im E_k$. In \fig{fig:unstablemodes} we show $E_k^2$ for several values of the spiral wave vector $Q$ for a one-dimensional Heisenberg spin system. Unstable modes are thus indicated by $E_k^2<0$, which for spirals with $Q<\pi/2a_\textrm{lat}$ appear at low wave vectors $k<Q$, while for spirals with $Q>\pi/2a_\textrm{lat}$ they occur at high wave vectors $Q>k$.
\begin{figure}[b]
        \centering
        \includegraphics[width=0.5\textwidth]{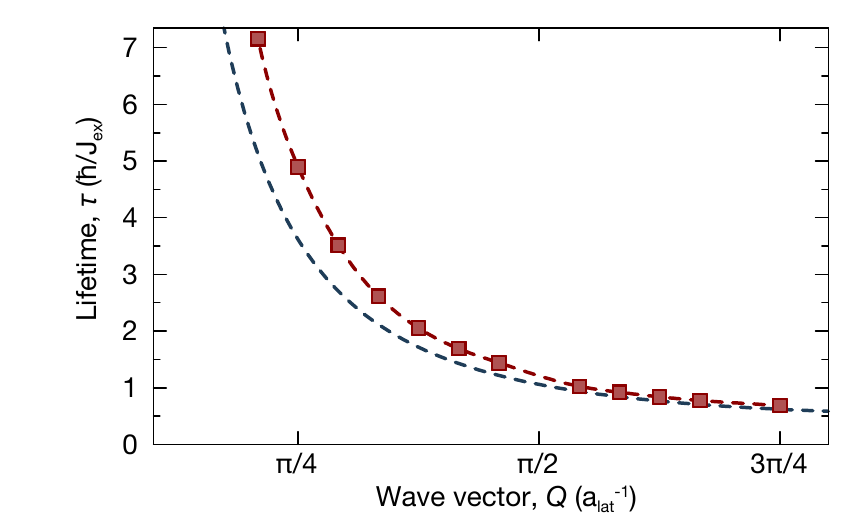}
        \caption{ Comparison of the spin spiral lifetime in one dimension obtained from numerically exact calculations (red squares) and from spin-wave theory (grey dashed line) for various wave vectors $Q$. }
        \label{fig:swted}
\end{figure}

\textbf{Lifetime of the spin spiral.---}Experimentally we extract the lifetime of the spiral through the decay of the $g_2$ correlation function which is equivalent to $ \langle \hat S_k^+ \hat S_k^-\rangle$. To evaluate this quantity we set up the equations of motion for the HP bosons in reciprocal space
\begin{align*}
 -i \frac{d}{dt} \langle a_k a_{-k} \rangle &= -2 \epsilon_k \langle a_k a_{-k}\rangle + \Delta_k (\langle a_k^\dag a_k^\nag\rangle + \langle a_{-k}^\dag a_{-k}^\nag\rangle)\\
 -i \frac{d}{dt} \langle a_k^\dag a_{k}^\nag \rangle &= \Delta_k (\langle a_k^\dag a_{-k}^\dag\rangle+\langle a_k a_{-k}\rangle)\;.
\end{align*}
This set of differential equations can be solved analytically. From the analytic solution we calculate the correlation function
\begin{align}
 \langle S_k^+ S_k^- \rangle &= \frac{1}{N} \sum_{i,j} e^{-ik (x_i-x_j)}\langle S_i^+ S_j^- \rangle
 =\frac{1}{4 S_0}\left(S_0 -  \langle a_{k-Q}^\dag a_{k-Q}^\nag \rangle+ \frac{1}{2} \right)\;.
\end{align}
In one dimension the lifetime of the spiral obtained within the spin-wave theory agrees reasonably well with the results obtained with numerically exact calculations, see \fig{fig:swted}. The numerical results are obtained with Time-Evolving Block Decimation (TEBD) for systems with open boundary conditions, while the spin-wave theory is calculated for systems with periodic boundary conditions. In higher dimension, we use spin-wave theory to predict the lifetime of the spiral. Furthermore, for $k=Q$, which is the only initially occupied mode, we find $\langle S_Q^+ S_Q^- \rangle \propto n_0 +\text{const}$ from which we determine the decay rate $\Gamma \propto \epsilon_0 \propto \sum_\alpha (1-\cos Q_\alpha) \underset{Q\to 0}{\sim} Q^2$, yielding the same low momentum scaling as the exact numerical results.

\section{Many-body spectrum}

The initial spin spiral evolves in time with the unitary quantum-mechanical time evolution operator
\begin{equation}
 \ket{\sp}(t) = e^{-i \hat H t}\ket{\sp} = \sum_\nu e^{-i  E_\nu t} \ket{\nu}\braket{\nu|\sp} \;.
\end{equation}
In the second step we inserted a resolution of identity spanned by the eigenstates of $\hat H$.
The dephasing dynamics is thus approximately given by the spread of the energies $\Delta E_\nu$ weighted by the
overlap of the corresponding eigenstate and the spin spiral $\braket{\nu|\sp}$. The decay rate
$\Gamma$ of the spiral prepared with a certain wave vector $Q$ should therefore also
be determined by this spread of energies $\Delta E_\nu$, which is discussed in \fig{fig:spectrum} of the main text. In particular, we analyze the full many-body spectrum of the ferromagnetic Heisenberg spin chain consisting of either 12 or 16 spins. We find a quadratic dependence of the weighted spread of eigenenergies $\Delta E_\nu$ on the spiral wave vector $Q$, which supports the scaling of the decay rate with $Q^2$ and thus provides a microscopic interpretation of the far-from-equilibrium dynamics in the one-dimensional Heisenberg chain.

\end{document}